\documentclass[aps,pre,twocolumn,superscriptaddress,nofootinbib]{revtex4}

\usepackage{graphicx}
\usepackage{amsmath}
\usepackage{amssymb}
\usepackage{color}
\usepackage{mathrsfs}
\usepackage[colorlinks=true,linkcolor=blue,citecolor=blue]{hyperref}
\usepackage[T1]{fontenc}

\newcommand{\be}{\begin{equation}}

\begin{document}

\title{Anomalous diffusion, non-Gaussianity, and nonergodicity for
subordinated fractional Brownian motion with a drift}

\author{Yingjie Liang}
\email{liangyj@hhu.edu.cn}
\affiliation{College of Mechanics and Materials, Hohai University,
211100 Nanjing, China}
\affiliation{University of Potsdam, Institute of Physics \& Astronomy,
14476 Potsdam-Golm, Germany}
\author{Wei Wang}
\email{weiwangnuaa@gmail.com}
\affiliation{University of Potsdam, Institute of Physics \& Astronomy,
14476 Potsdam-Golm, Germany}
\author{Ralf Metzler}
\email{rmetzler@uni-potsdam.de}
\affiliation{University of Potsdam, Institute of Physics \& Astronomy,
14476 Potsdam-Golm, Germany}
\affiliation{Asia Pacific Centre for Theoretical Physics, Pohang 37673,
Republic of Korea}

\date{\today}

\begin{abstract}
The stochastic motion of a particle with long-range correlated increments
(the moving phase) which is intermittently interrupted by immobilizations
(the traping phase) in a disordered medium is considered in the presence
of an external drift. In particular, we consider trapping events whose
times follow a scale-free distribution with diverging mean trapping time.
We construct this process in terms of fractional Brownian motion (FBM)
with constant forcing in which the trapping effect is introduced by the
subordination technique, connecting "operational time" with observable
"real time". We derive the statistical properties of this process such as
non-Gaussianity and non-ergodicity, for both ensemble and single-trajectory
(time) averages. We demonstrate nice agreement with extensive simulations for
the probability density function, skewness, kurtosis, as well as ensemble
and time-averaged mean squared displacements. We pay specific emphasis on
the comparisons between the cases with and without drift.
\end{abstract}

\maketitle

\section{Introduction}
\label{sec-1}

Brownian motion (normal diffusion) is characterized by a mean squared
displacement (MSD) of a tracer particle that is a linear function of time
\cite{vankampen,haenggirev,spiecho,levybrown}. Moreover, the probability
density function (PDF) of the displacements is Gaussian. The emergence of
such normal diffusion rests on the following three conditions: (i) there
exists a finite correlation time after which individual displacements
become independent, (ii) the displacements are identically distributed, and
(iii) the second moment of the displacements is finite. Anomalous diffusion,
in contradistinction, which may appear whenever one of these conditions is
violated, has been widely observed in systems ranging from soft and bio
matter over condensed matter, up to financial markets or geophysical systems
\cite{bouc,magi13,sun,bols,soko,metz00,metz14}. In anomalous diffusion the MSD
follows the power-law form
\begin{equation}
\label{msd}
\langle x^2(t)\rangle\sim K_{\alpha}t^{\alpha},
\end{equation}
where the anomalous diffusion coefficient has physical dimension $[K_{\alpha}]
=\mathrm{length}^2/\mathrm{time}^{\alpha}$. Depending on the value of the
anomalous diffusion exponent we distinguish subdiffusion for $0<\alpha<1$
from superdiffusion for $\alpha>1$ \cite{bouc,metz00,soko,bark12}. Examples
for subdiffusion include the motion of submicron tracers in the crowded
environment of living biological cells \cite{golding,weber,lene} or in
polymer-crowded in vitro liquids \cite{szym,lene1}, the motion of
potassium channels resident in the plasma membranes of living cells
\cite{bark12,pt1}. For superdiffusion we mention motor-driven transport
of viruses \cite{seis}, neuronal messenger ribonucleoprotein \cite{song},
endogeneous cellular vesicels \cite{christine} or magnetic endosomes
\cite{robe10}.

As soon as one or more of the above three conditions for normal diffusion 
are violated, the resulting stochastic process is no longer universal in
the sense that a given measured MSD (\ref{msd}) may result from different
processes \cite{soko,bark12,pt1,metz14}. The inference of such processes
from data has received considerable attention, approaches including the
construction of decision trees based on complementary statistical
observables \cite{yazmine}, dynamic scaling exponents \cite{philipp,vilk1},
or p-variations \cite{marcin}, as well as "objective" methods such as
Bayesian analysis \cite{michael,samu,samu1} or machine learning
\cite{yael,gorka,janusz,janusz1,bo,andi,henrik}, inter alia.

Here we consider the case when two specific, fundamental anomalous
stochastic processes are combined in the presence of an external drift. These
two processes are fractional Brownian motion (FBM) and the subdiffusive
continuous time random walk (CTRW). Going back to Kolmogorov \cite{kolmo} and
Mandelbrot and van Ness \cite{mand}, FBM is a non-Markovian process driven by
zero-mean, stationary Gaussian noise with long-range correlations. FBM can
describe both sub- and superdiffusion, depending on whether the noise is
antipersistent or persistent (see below). Examples for subdiffusive FBM-type
motion include tracer motion in complex liquids \cite{szym,lene1} and in the
cytoplasm of living cells \cite{lene,weber}, or lipid dynamics in bilayer
membranes \cite{jeon}. Superdiffusive FBM was found for the motion of higher
animals \cite{vilk1} and of vacuoles in amoeboid cells and the amoeba cells
themselves \cite{krap}, and it is used as a model for densities of persistently
growing brain fibers \cite{janu20}.

The continuous time random walk (CTRW) model introduced by Montroll and Weiss
\cite{mont} is a generalization of a random walk in which the particle waits
for a random time $\tau$ between jumps. When the associated PDF of waiting
times is scale free, $\psi(\tau)\simeq\tau^{-1-\alpha}$ with $0<\alpha<1$,
such that the mean waiting time $\langle\tau\rangle$ diverges, subdiffusion
of the form (\ref{msd}) emerges, where the anomalous diffusion exponent is
given by the scaling exponent of the waiting time PDF. Examples for such
CTRW subdiffusion include the motion of charge carriers in amorphous
semiconductors \cite{scher,neher}, and (asymptotic) power-law forms for
the waiting times $\tau$ were identified, i.a., for the motion of potassium
channels in live cell membranes \cite{weig}, in glass-forming liquids
\cite{rubn}, for drug molecules diffusing in between two silica slabs
\cite{amanda}, and for tracer transport in geological formations \cite{berk}.

There exist a growing number of examples in which analysis of the recorded
dynamics demonstrates the conspirative action of more than a single anomalous
diffusion process. In particular, we here mention systems in which CTRW and
FBM simultaneously. These include the motion of insulin granules in living
MIN6 insulisoma cells \cite{tabei}, of nicotinic acetylcholine membrane
receptors \cite{mosqueira}, nanosized tracer objects in the cytoplasm
\cite{etoc}, drug molecules confined by silica slabs \cite{amanda}, and of
voltage-gated sodium channels on the surface of hippocampal neurons \cite{fox}.

The combined stochastic process of long-range correlated FBM and scale-free
waiting time effects was recently studied in terms of a subordination concept
in \cite{fox}. We here go one step beyond and study this type of stochastic
motion in the presence of a constant external drift. To this end we start our
description with an FBM process, which is then subordinated to a stable
subordinator \cite{appl,boch,fell,chec21,meer13,meer04,magd,wang22}.
Historically, the notion of subordination was introduced by Bochner
\cite{boch} and applied by Feller \cite{fell}. The dynamics of the process
of interest here thus involves three elements, namely, long-range correlations,
scale-free waiting times, and an external drift. We here investigate
analyitcally and numerically the transport properties of this process.

The paper is organized as follows. In Sec.~\ref{sec-2}, the Langevin equation
for subordinated FBM in the presence of a drift is introduced. Sec.~\ref{sec-3}
provides the PDF and shows its non-Gaussianity. We also obtain the
ensemble-averaged MSD. In Sec.~\ref{sec-4} the time-averaged MSD (TAMSD) is
derived, discussing the weakly non-ergodic statistic along with the amplitude
scatter of the TAMSD. The results are discussed in Sec. \ref{sec-6}. For the
convenience of the reader the results for subordinated FBM without drift are
briefly summarized in App.~\ref{app-a}.

\section{Subordinated fractional Brownian motion with drift}
\label{sec-2}

Subordinated FBM with drift, $x(t)\equiv x(s(t))$ combines both long trapping
times and long-range correlations with a drift. It satisfies the coupled
stochastic equations
\begin{eqnarray}
\nonumber
\frac{dx(s)}{ds}&=&v+\sqrt{2D}\zeta_H(s),\\
\frac{dt(s)}{ds}&=&\varepsilon(s),
\label{eq-0}
\end{eqnarray}
see the discussions in \cite{fogedby,magd}. Here $x(s)$ is the particle
trajectory as function of "operational time" $s$, and $v$ is the constant
external drift. Without loss of generality we set $D=0.5$. Moreover, $\zeta
_H(s)$ represents fractional Gaussian noise with correlation function
$\langle\zeta_H(t_1)\zeta_H(t_2)\rangle\sim H(2H-1)|t_1-t_2|^{2H-2}$,
for $t_1\neq t_2$ \cite{mand}. Here, $H$ with $0<H<1$ is the Hurst exponent.
For free FBM, this effects the MSD (\ref{msd}) with $\alpha=2H$. The second
equation then translates the
operational time $s$ into the "real" process time $t$, where $\varepsilon(s)$
represents one-sided L{\'e}vy stable noise \cite{gawr,pens10}, which is the
formal derivative of the L{\'e}vy stable subordinator $t(s)$ with stability
index $0<\alpha<1$. The L{\'e}vy stable subordinator is a non-decreasing
L{\'e}vy process with stationary and independent increments. The one-sided
L{\'e}vy stable distribution \cite{saa,pens16,datt} is defined in terms of
its Laplace transform via $\hat{L}_\alpha(k)=\exp(-k^\alpha)$, which is
strictly increasing. The inverse subordinator $s(t)$ is defined as $s(t)=
\inf\{s>0:t(s)>t\}$, where $s(t)$ is called the hitting time or first-passage
time process \cite{meer13}, which can be considered as the limit process of
the continuous time random walk with a heavy tailed waiting time PDF. It tends
to infinity when $t\to\infty$. We note that the distribution of $s(t)$ is also
called the Mittag-Leffler distribution based on the relationship between the
moments and the Mittag-Leffler function \cite{saa}. The subordinator $s(t)$
is responsible for the subdiffusive behavior with long rests of the particle,
while the parent process $x(s)$ introduces the correlated FBM with drift.

\begin{figure}
\includegraphics[width=8cm]{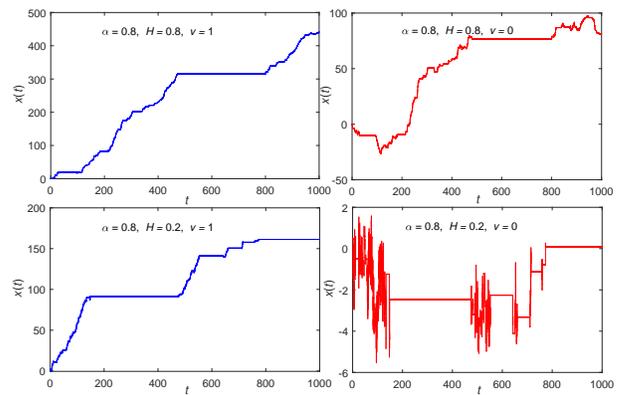}
\caption{Sample trajectories of subordinated FBM $x(t)$ with drift $v=1$ and
without drift ($v=0$) for different values of the Hurst exponent $H$ (see the
legends in the panels) of the parental FBM and waiting time index $\alpha=0.8$.
Note the different spans of the vertical axes. The elementary time step in the
simulations was $dt=0.1$.}
\label{fig-1}
\end{figure}

Fig.~\ref{fig-1} shows sample trajectories of subordinated FBM with drift
$v=1$ and without drift ($v=0$) for two values of the Hurst index (persisent,
i.e., positively correlated FBM, with $H=0.8$ and anti-persistent, negatively
correlated FBM with $H=0.2$), and for $\alpha=0.8$. We use identical noise time
series of the fractional Gaussian noise time series and the subordinator for
each $H$ value. The cases for $\alpha=1$, i.e., without long waiting times,
are shown in Fig.~\ref{fig-a5} in App.~\ref{app-a}. Even for the relatively
large waiting time exponent $\alpha=0.8$ the effects of immobilization are
clearly present.

\section{Probability density function, moments, and ensemble-averaged mean
squared displacement}
\label{sec-3}

The PDF of the subordinated process is \cite{bark} 
\begin{equation}
\label{eq-1}
P(x,t)=\int_0^{+\infty}P_0(x,s)h(s,t)ds, 
\end{equation} 
where $P_0(x,s)$ and $h(s,t)$ denote the PDFs of the parental process $x(s)$
and of the inverse stable subordinator $s(t)$, respectively. Specifically,
the PDF of the original process $x(s)$ is
\begin{equation}
\label{eq-2}
P_0(x,s)=\frac{1}{\sqrt{2\pi s^{2H}}}\exp\left(-\frac{(x-vs)^2}{2s^{2H}}\right),
\end{equation}
and the PDF of the inverse stable subordinator $s(t)$ is \cite{bark}
\begin{equation}
\label{eq-3}
h(s,t)=\frac{t}{\alpha s^{1+1/\alpha}}L_\alpha\left(t/s^{1/\alpha}\right),
\end{equation}
in terms of the one-sided L{\'e}vy stable distribution $L_{\alpha}$. Then
Eq.~\eqref{eq-1} can be rewritten as 
\begin{eqnarray}
\nonumber
P(x,t)&=&\int_0^{\infty}\frac{1}{\sqrt{2\pi s^{2 H}}}\exp\left(-\frac{(x-vs)
^2}{2s^{2H}}\right)\\
&&\times\frac{t}{\alpha s^{1+1/\alpha}}L_\alpha\left(\frac{t}{s^{1/\alpha}}
\right)ds. 
\label{eq-4}
\end{eqnarray}
The moments $\langle x^n(t)\rangle$ of $x(t)$ are \cite{pens16}
\begin{equation}
\label{eq-5}
\langle x^n(t)\rangle=\int_0^{\infty}\langle x^n(s)\rangle h(s,t)ds,
\end{equation}
where $\langle x^n(s)\rangle$ represent the moments of the parental FBM process
with drift. The first moment, $\mu$, is then
\begin{equation}
\label{eq-6}
\mu=\langle x(t)\rangle=\int_0^{\infty}vsh(s,t)ds=\frac{v}{\Gamma(1+\alpha)}
t^\alpha.
\end{equation}
The second moment $\langle x^2(t)\rangle$ reads
\begin{eqnarray}
\nonumber
\langle x^2(t)\rangle&=&\int_0^{\infty}\left(s^{2 H}+v^2s^2\right)h(s,t)d\tau\\
&=&\frac{\Gamma(1+2H)}{\Gamma(1+2H\alpha)}t^{2H\alpha}+\frac{2v^2}{\Gamma(1+2
\alpha)}t^{2\alpha},
\label{eq-7}
\end{eqnarray}
in which asymptotically the drift term will be dominant ($H<1$). The MSD (or
variance) corresponds to the second central moment 
\begin{eqnarray}
\nonumber
\sigma^2&=&\langle\Delta x^2(t)\rangle\\
\nonumber
&=&\frac{\Gamma(1+2H)}{\Gamma(1+2H\alpha)}t^{2H\alpha}\\
&&+\left(\frac{2}{\Gamma(1+2\alpha)}-\frac{1}{\Gamma(1+\alpha)^2}\right)v^2
t^{2\alpha},
\label{eq-8}
\end{eqnarray}
where $\sigma$ is the standard derivation.

The coefficient of variation $c=\sigma/\mu$ for this motion is given by
\begin{equation}
\label{eq-9}
c_v=\frac{\left(\frac{\Gamma(1+2H)}{\Gamma(1+2H\alpha)}t^{2H\alpha}+\left(
\frac{2}{\Gamma(1+2\alpha)}-\frac{1}{\Gamma(1+\alpha)^2}\right)v^2t^{2\alpha}
\right)^{1/2}}{\frac{v}{\Gamma(1+\alpha)}t^\alpha},
\end{equation}
and the skewness $\theta=\langle[(x-\mu)/\sigma]^3\rangle$ becomes
\begin{equation}
\label{eq-10}
\theta=\frac{\langle(x(t)-\mu)^3\rangle}{\langle(x(t)-\mu)^2\rangle^{3/2}},
\end{equation}
for which we need to obtain the third central moment and the third power of
the standard variation,
\begin{widetext}
\begin{eqnarray}
\nonumber
\langle(x(t)-\mu)^3\rangle&=&\left(\frac{3\Gamma(2+2H)v}{\Gamma(1+\alpha+2H
\alpha)}-\frac{3\Gamma(2+2H)v}{(1+2H)\Gamma(1+2H\alpha)\Gamma(1+\alpha)}
\right)t^{2H\alpha+2\alpha}\\
&&+\left(\frac{6v^3}{\Gamma(1+3\alpha)}-\frac{6v^3}{\Gamma(1+2\alpha)\Gamma(
1+\alpha)}+\frac{2v^3}{\Gamma(1+\alpha)^3}\right)t^{3 \alpha},
\label{eq-11}
\end{eqnarray}
and $\langle(x(t)-\mu)^2\rangle^{3/2}=\left(\sigma^2\right)^{3/2}$ can be
calculated using Eq.~\eqref{eq-8}. The skewness is 0, i.e., the PDF is symmetric
when the drift vanishes ($v=0$).

The kurtosis $\kappa=\langle[(x-\mu)/\sigma]^4
\rangle=\langle(x(t)-\mu)^4\rangle/\langle(x(t)-\mu)^2\rangle^2$ can be
calculated based from the fourth central moment and the fourth power of the
standard deviation,
\begin{eqnarray}
\nonumber
\langle(x(t)-\mu)^4\rangle&=&\frac{3\Gamma(1+4H)}{\Gamma(1+4H\alpha)}t^{4H\alpha}
+\left(\frac{6\Gamma(3+2H)v^2}{\Gamma(1+2H\alpha+2\alpha)}-\frac{12\Gamma(3+2H)
v^2}{(2+2H)\Gamma(1+2H\alpha+\alpha)\Gamma(1+\alpha)}\right.\\
\nonumber
&&\left.+\frac{6\Gamma(3+2H)v^2}{(1+2H)(2+2H)\Gamma(1+2H\alpha)\Gamma(1+\alpha)
^2}\right)t^{2H\alpha+2\alpha}\\
&&+\left(\frac{24v^4}{\Gamma(1+4\alpha)}-\frac{24v^4}{\Gamma(1+3\alpha)\Gamma(
1+\alpha)}+\frac{12v^4}{\Gamma(1+2\alpha)\Gamma(1+\alpha)^2}-\frac{3 v^4}{\Gamma
(1+\alpha)^4}\right)t^{4\alpha},
\label{eq-13}
\end{eqnarray}
and
\begin{eqnarray}
\nonumber
\langle(x(t)-\mu)^2\rangle^2&=&\frac{\Gamma(1+2H)^2}{\Gamma(1+2H\alpha)^2}t^{4H
\alpha}+\left(\frac{4\Gamma(1+2H)v^2}{\Gamma(1+2H\alpha)\Gamma(1+2\alpha)}-
\frac{2\Gamma(1+2H)v^2}{\Gamma(1+2H\alpha)\Gamma(1+\alpha)^2}\right)t^{2H
\alpha+2\alpha}\\
&&+\left(\frac{4v^4}{\Gamma(1+2\alpha)^2}-\frac{4v^4}{\Gamma(1+2\alpha)\Gamma(
1+\alpha)^2}+\frac{v^4}{\Gamma(1+\alpha)^4}\right)t^{4\alpha}.
\label{eq-14}
\end{eqnarray}
When $t\to\infty$ the kurtosis will be time and drift independent, and determined
by the stability index,
\begin{equation}
\label{eq-15}
\kappa=\frac{\displaystyle\frac{24}{\Gamma(1+4\alpha)}-\frac{24}{\Gamma(1+3
\alpha)\Gamma(1+\alpha)}+\frac{12}{\Gamma(1+2\alpha)\Gamma(1+\alpha)^2}-\frac{3}{\Gamma(1+\alpha)^4}}{\displaystyle\frac{4}{\Gamma(1+2 \alpha)^2}-
\frac{4}{\Gamma(1+2\alpha)\Gamma(1+\alpha)^2}+\frac{1}{\Gamma(1+\alpha)^4}}.
\end{equation}
\end{widetext}

\begin{figure}
\includegraphics[width=8cm]{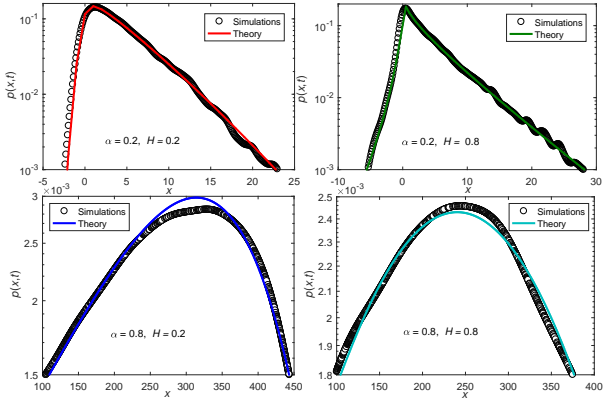}
\caption{Simulations (dark circles) and analytical results (solid curves) from
Eq.~\eqref{eq-4} for the PDF of subordinated FBM with drift $v=1$ at time $t=1000$
for varying Hurst exponents $H$ and stability indices $\alpha$. The elementary
time step in the simulations was $dt=0.1$.}
\label{fig-2} \end{figure}

\begin{figure}
\includegraphics[width=8cm]{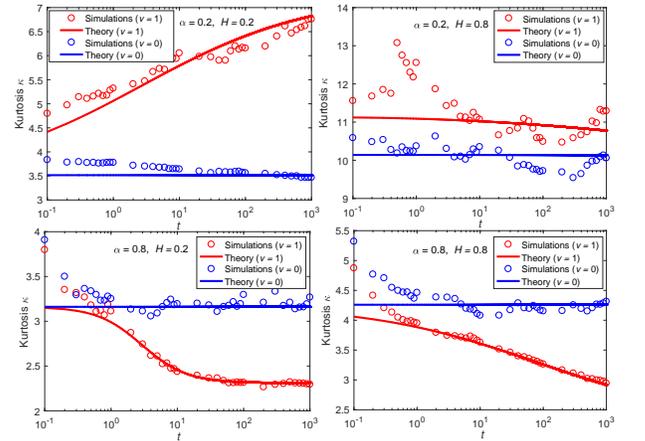}
\caption{Simulations (circles) and theoretical results (solid curves) based
on Eqs. \eqref{eq-13} and \eqref{eq-14} of the kurtosis for the subordinated
FBM with the drift $v = 1$ and without drift $v = 0$ for varying values
of Hurst exponent $H$ and stability index $\alpha$ with increasing time.
The elementary time step in the simulations was $dt=0.1$.}
\label{fig-3}
\end{figure}

In Fig.~\ref{fig-2} we show the results of our analytical calculations and
stochastic simulations of the PDF with drift $v=1$ at time $t=1000$ for the
Hurst exponents $H=0.2$ and $H=0.8$, and stability indices $\alpha=0.2$ and
$\alpha=0.8$. The analytical results agree nicely with the simulations for
all cases, apart from deviations aorund the maxima for the case $\alpha=0.8$.
The PDFs are asymmetric as compared with the symmetric PDFs of subordinated
FBM without drift in Fig.~\ref{fig-a1} (see App.~\ref{app-a}). All PDFs have
distint cusp-like peaks for smaller $\alpha$ both in the presence and absence
of drift. The PDFs of subordinated FBM with $\alpha=1$ (i.e., the operational
time has the same mean behavior as the process time) are shown in
Fig.~\ref{fig-a6} (App.~\ref{app-a}).

To check the non-Gaussianity of the PDF for subordinated FBM with drift, we
illustrate in Fig.~\ref{fig-3} the kurtosis based on Eqs.~\eqref{eq-13} and
\eqref{eq-14} with the drift $v=1$ and without drift ($v=0$) from analytical
calculations and simulations, as function of time. For different values of
$H$ and $\alpha$ the general agreement is good. However, due to the fourth
order of the means entering the kurtosis, we did not manage to achieve a
higher numerical accuracy from our simulations. The results for the kurtosis,
whose value for a Gaussian in one dimension is $\kappa=3$, indicate
that the PDFs are non-Gaussian, which is in full agreement with the results
given in Figs.~\ref{fig-2} and \ref{fig-a1} (App.~\ref{app-a}). We note
specifically that for smaller values of $\alpha$ or larger value of $H$, the
values of the kurtosis exhibit values that are much larger than the Gaussian
value 3.

\section{Time averaged mean squared displacement and its distribution}
\label{sec-4}

The TAMSD is defined as \cite{he,metz14,pt1,bark12} 
\begin{equation}
\label{eq-16}
\overline{\delta^2(\Delta)}=\frac{1}{T-\Delta}\int_0^{T-\Delta}[x(t+\Delta)
-x(t)]^2dt,
\end{equation}
where $T$ is the measurement time and $\Delta$ is called the lag time. The
average TAMSD is based on the autocorrelation function $\langle x(t_1)x(t_2)
\rangle$, which is given by
\begin{widetext}
\begin{eqnarray}
\nonumber
\langle x(t_1)x(t_2)\rangle&=&\int_0^{\infty}\int_0^{\infty}\langle x(s_1)
x(s_2)\rangle h(s_2,t_2,s_1,t_1)ds_1ds_2\\
\nonumber
&=&\int_0^{\infty}\int_0^{\infty}\left[v^2s_1s_2+\frac{1}{2}\left(s_1^{2H}+
s_2^{2H}-|s_2-s_1|^{2H}\right)\right]h(s_2,t_2;s_1,t_1)ds_1ds_2\\
&=&v^2\langle s(t_1)s(t_2)\rangle+\frac{1}{2}\left[\langle s(t_1)^{2 H}\rangle
+\langle s(t_2)^{2H}\rangle-\langle|s(t_2)-s(t_1)|^{2H}\rangle\right].
\label{eq-17}
\end{eqnarray}

As we argued in Sec.~\ref{sec-2}, $s(t)$ is the hitting time, also called the
number of steps up to time $t$. Therefore
\begin{equation}
\label{eq-18} 
\langle[x(t_2)-x(t_1)]^2\rangle=\langle x^2(t_2)\rangle+\langle x^2(t_1)
\rangle-2\langle x(t_1)x(t_2)\rangle=v^2\langle[s(t_2)-s(t_1)]^2\rangle
+\langle|s(t_2)-s(t_1)|^{2H}\rangle.
\end{equation}
Let $t_1=t$ and $t_2=t+\Delta$, then the fractional moment of order $\vartheta$
of $s(t_2)-s(t_1)$ is \cite{fox}
\begin{equation}
\label{eq-19}
\langle[s(t+\Delta)-s(t)]^{\vartheta}\rangle=\frac{\Gamma(1+\vartheta)}{
\Gamma(\alpha)\Gamma(2-\alpha+\alpha\vartheta)}\,_2F_1\left(1,1-\alpha;2
-\alpha+\alpha\vartheta;-\frac{\Delta}{t}\right)\frac{\Delta^{1-\alpha
+\alpha\vartheta}}{t^{1-\alpha}},
\end{equation}
where $_2F_1$ is the hypergeometric function. With the help of Eqs.~\eqref{eq-16}
and \eqref{eq-19} we obtain the mean TAMSD in the form
\begin{eqnarray}
\nonumber
\left<\overline{\delta^2(\Delta)}\right>&=&\frac{1}{T-\Delta}\int_0^{T-\Delta}
\left(\frac{2v^2}{\Gamma(\alpha)\Gamma(2+\alpha)}\,_2 F_1\left(1,1-\alpha;2+
\alpha;-\frac{\Delta}{t}\right)\frac{\Delta^{1+\alpha}}{t^{1-\alpha}}\right.\\
&&\left.+\frac{\Gamma(1+2H)}{\Gamma(\alpha)\Gamma(2-\alpha+2H\alpha)}\,_2 F_1
\left(1,1-\alpha;2-\alpha+2H\alpha;-\frac{\Delta}{t}\right)\frac{\Delta^{1-
\alpha+2H\alpha}}{t^{1-\alpha}}\right)dt.
\label{eq-20}
\end{eqnarray}

In the limit $\Delta\ll t$ we have $_2F_1(1,1-\alpha;2-\alpha+\alpha\vartheta;0)
\sim1$ \cite{virch}, then Eq.~\eqref{eq-20} becomes
\begin{equation}
\label{eq-21}
\left<[x(t_2)-x(t_1)]^2\right>\sim\frac{2}{\Gamma(\alpha)\Gamma(2+\alpha)}\frac{
v^2\Delta^{1+\alpha}}{t^{1-\alpha}}+\frac{\Gamma(1+2H)}{\Gamma(\alpha)\Gamma(
2-\alpha+2H\alpha)}\frac{\Delta^{1-\alpha+2H\alpha}}{t^{1-\alpha}}.
\end{equation}
Performing the temporal integration in Eq.~\eqref{eq-16} we get
\begin{equation}
\label{eq-22}
\left<\overline{\delta^2(\Delta)}\right>\sim\frac{2v^2}{\alpha\Gamma(\alpha)
\Gamma(2+\alpha)}\frac{\Delta^{1+\alpha}}{T^{1-\alpha}}+\frac{\Gamma(1+2H)}{
\alpha\Gamma(\alpha)\Gamma(2-\alpha+2H\alpha)}\frac{\Delta^{1-\alpha+2H
\alpha}}{T^{1-\alpha}}.
\end{equation}
In the limit $\Delta\gg t$, we have $_2F_1\left(1,1-\alpha;2-\alpha+\alpha
\vartheta;-\frac{\Delta}{t}\right)\sim\left(\frac{t+\Delta}{t}\right)^{-(
1-\alpha)}\frac{\Gamma(2-\alpha+\alpha\vartheta)\Gamma(\alpha)}{\Gamma(1+
\alpha \vartheta)}$, and Eq.~\eqref{eq-20} becomes
\begin{equation}
\label{eq-23}
\left<[x(t_2)-x(t_1)]^2\right>\sim\frac{2v^2}{\Gamma(1+2\alpha)}\left(\Delta
^2+(\alpha-1)t\Delta^{2\alpha-1}\right)+\frac{\Gamma(1+2 H)}{\Gamma(1+2H
\alpha)}\left(\Delta^{2H\alpha}+(\alpha-1)t\Delta^{2H\alpha-1}\right).
\end{equation}
Similarly from Eq. \eqref{eq-16} we get
\begin{equation}
\label{eq-24}
\left<\overline{\delta^2(\Delta)}\right>\sim\frac{2v^2}{\Gamma(1+2 \alpha)}\left(
\Delta^2+\frac{(\alpha-1)(T-\Delta)\Delta^{2\alpha-1}}{2}\right)+\frac{\Gamma(
1+2H)}{\Gamma(1+2H\alpha)}\left(\Delta^{2H\alpha}+\frac{(\alpha-1)(T-\Delta)
\Delta^{2H\alpha-1}}{2}\right).
\end{equation}
We note that subordinated FBM with drift for the limit $\alpha=1$ is an
ergodic stochastic process \cite{deng,schw,jae_pre12}.
\end{widetext}
 
\begin{figure}
\includegraphics[width=8cm]{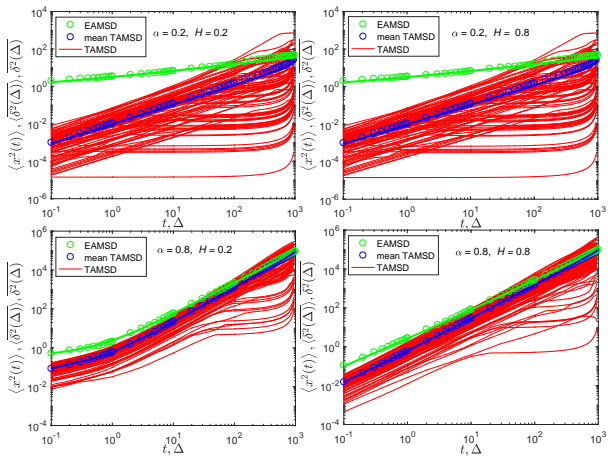}
\caption{MSD (green circles), a number of individual TAMSDs (thin red curves),
and the mean TAMSD (thick blue circles) for subordinated FBM with drift $v=1$
and different values of $H$ and $\alpha$. Theoretical results for the MSD
(thick solid green curves) and mean TAMSD (thick solid blue curves) as given
in Eqs.~\eqref{eq-7} and \eqref{eq-20}, respectively, show nice agreement.
Parameters: length of the trajectories $T=1000$, elementary time-step $dt=0.1$,
and number of trajectories $n=300$.}
\label{fig-4}
\end{figure}

Fig.~\ref{fig-4} shows results from simulations of the MSD, a number of
individual TAMSDs, and the mean TAMSD for subordinated FBM with drift
$v=1$ for different values of $H$ and $\alpha$. In Fig.~\ref{fig-4} the
analytical results are in nice agreement with the simulations for the MSD
and the mean TAMSD. We also note that individual TAMSDs show a wide spread,
especially for smaller values of $\alpha$, consistent with the weakly
non-ergodic behavior of CTRW motion \cite{he,bark12,pt1,metz14}. The values
of the MSD and TAMSD grow faster with time for subordinated FBM with drift
as compared to the same motion in absence of the drift, as shown in
Fig.~\ref{fig-a2} (App.~\ref{app-a}).

To quantify the relative amplitude spread of individual TAMSD, we use the
dimensionless variable \cite{he,bark12,metz14}
\begin{equation}
\label{eq-25}
\xi(\Delta)=\frac{\overline{\delta^2(\Delta)}}{\left<\overline{
\delta^2(\Delta)}\right>}.
\end{equation}
The values of $\xi$ fluctuate around the mean value $\left<\xi(\Delta)\right>
=1$. The corresponding PDF $\phi(\xi)$ can be expressed as modified totally
asymmetric L{\'e}vy stable density \cite{he,metz14} or via the Mittag-Leffler
function \cite{needcite}. The variance $\mathrm{EB}=\langle\xi^2\rangle-1$ is
called the ergodicity breaking parameter \cite{he,bark12,metz14,pt1}.

\begin{figure}
\includegraphics[width=8cm]{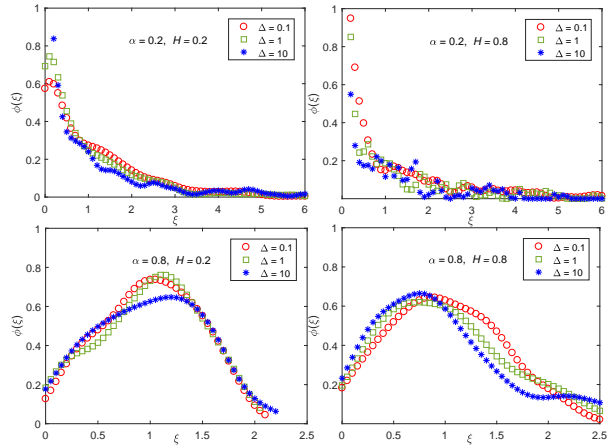}
\caption{Amplitude scatter PDF $\phi(\xi)$ of TAMSDs for subordinated FBM with
drift $v=1$ and varying values of $H$ and $\alpha$. The symbols (circle, square,
and star) respectively denote simulations with the lag time $\Delta=0.1$, $1$,
and $10$.}
\label{fig-5}
\end{figure}

Fig.~\ref{fig-5} illustrates the scatter PDF of the TAMSD for the same
parameters as in Fig.~\ref{fig-4} for the three different lag times $\Delta=
0.1$, $1$, and $10$. The results show that the distributions are not
symmetric around the mean $\left<\xi(\Delta)\right>=1$, and have a spike and
long heavy tail for smaller values of $\alpha$. It is also noted that for
smaller values of $\alpha$ the particles will have longer average waiting
times, and thus the probability will be higher for trajectories without
any displacement up to $T$, $\xi(\Delta)=0$, which agrees well with the
general observed trends in Fig.~\ref{fig-4} (see also the discussions in
\cite{he,metz14,fox}). The simulations of the scatter PDF for subordinated
FBM without drift have similar statistical properties, see Fig.~\ref{fig-a3}
(App.~\ref{app-a}).

\section{Discussion}
\label{sec-6}

We studied subordinated FBM in the presence of a constant external drift,
combining long-range correlated Gaussian motion characterized by a Hurst
exponent $H$ and long-tailed, scale-free PDFs of immobilization (waiting)
times with scaling exponent $\alpha$. As expected, at long times the drift
term dominates the transport behavior, which we quantified in terms of the
ensemble- and time-averaged moments as well as the PDF. Technically, we
employ the subordination approach based on a stable subordinator. This
transforms the "operational time" of the parental FBM to the "process time"
of the combined motion in the presence of the immobilization events. The
resulting process, studied in \cite{fox} in the absence of drift, thus
combines two central properties of stochastic motion observed in a wide
range of experiments. Currently, such observations predominantly come from
single-particle tracking in soft- and bio-matter \cite{manz,jaqa,etoc} or
large-scale computer simulations, see, e.g., \cite{amanda,jeon}. Given the
development of experimental methods to record single-particle movement in
geophysical contexts \cite{hard,roub}, it will be interesting to see whether 
similarly rich behaviors are unveiled in this context.

In our analysis we highlighted the different scaling behaviors in the MSDs
due to diffusion and drift, respectively. The resulting PDF is non-Gaussian,
and we showed how $H$ and $\alpha$ influence the shape parameters (skewness
and kurtosis). We also studied from simulations the amplitude scatter of
individual TAMSDs. In future work we will also consider aging effects, for
which explicit expressions for the ensemble-and time-averaged moments will
be obtained.

Subordinated FBM as studied in \cite{fox} in absence of a drift and with a
drift as investigated herein, complements similar combined stochastic
processes reported in literature. We here mention the combination of
CTRWs on fractal structures such as a Sierpi{\'n}ski gasket \cite{mero},
the conspiracy of FBM with scaled Brownian motion (SBM) \cite{andreywei} in
which the diffusion coefficient is a power-law function of time \cite{lim,jae},
and the combination of FBM with a stochastically evolving ("diffusing")
diffusivity \cite{wei,wei1}. Such processes will form the basis for future
extensions of data analyses based on statistical observables (such as those
developed herein), Bayesian maximum likelihood approaches \cite{michael,samu,
samu1}, or machine-learning strategies \cite{janusz,janusz1,bo,yael,gorka,andi,
henrik}.

\begin{acknowledgments}
Y. L. acknowledges financial support from the Alexander von Humboldt Foundation
(grant no. 1217531) and the National Natural Science Foundation of China
(grant no. 11702085). R. M. acknowledges financial support from the German
Science Foundation (DFG, grant no. ME 1535/12-1).
\end{acknowledgments}

\appendix

\numberwithin{figure}{section} 
\renewcommand{\thefigure}{S\arabic{figure}}

\section{Supplementary figures}
\label{app-a}

We here present additional figures complementing the behaviors presented
in the main text.

\begin{figure}
\includegraphics[width=8cm]{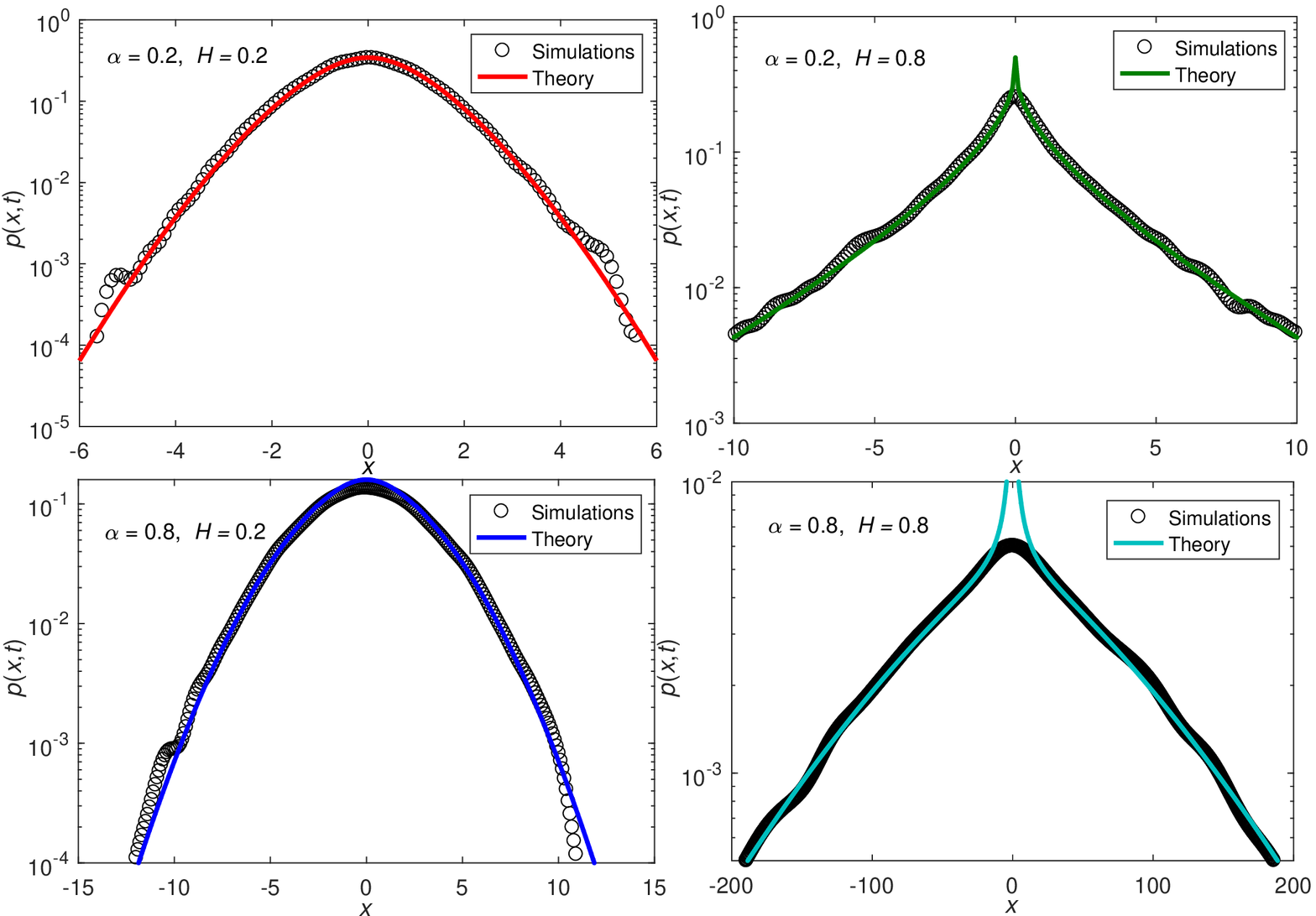}
\caption{Simulations (dark circles) and analytical results (solid curves)
from Eq.~\eqref{eq-5} for the PDF of subordinated FBM without drift ($v=0$)
and for different $H$ and $\alpha$, for $t=1000$.}
\label{fig-a1}
\end{figure}

\begin{figure}
\includegraphics[width=8cm]{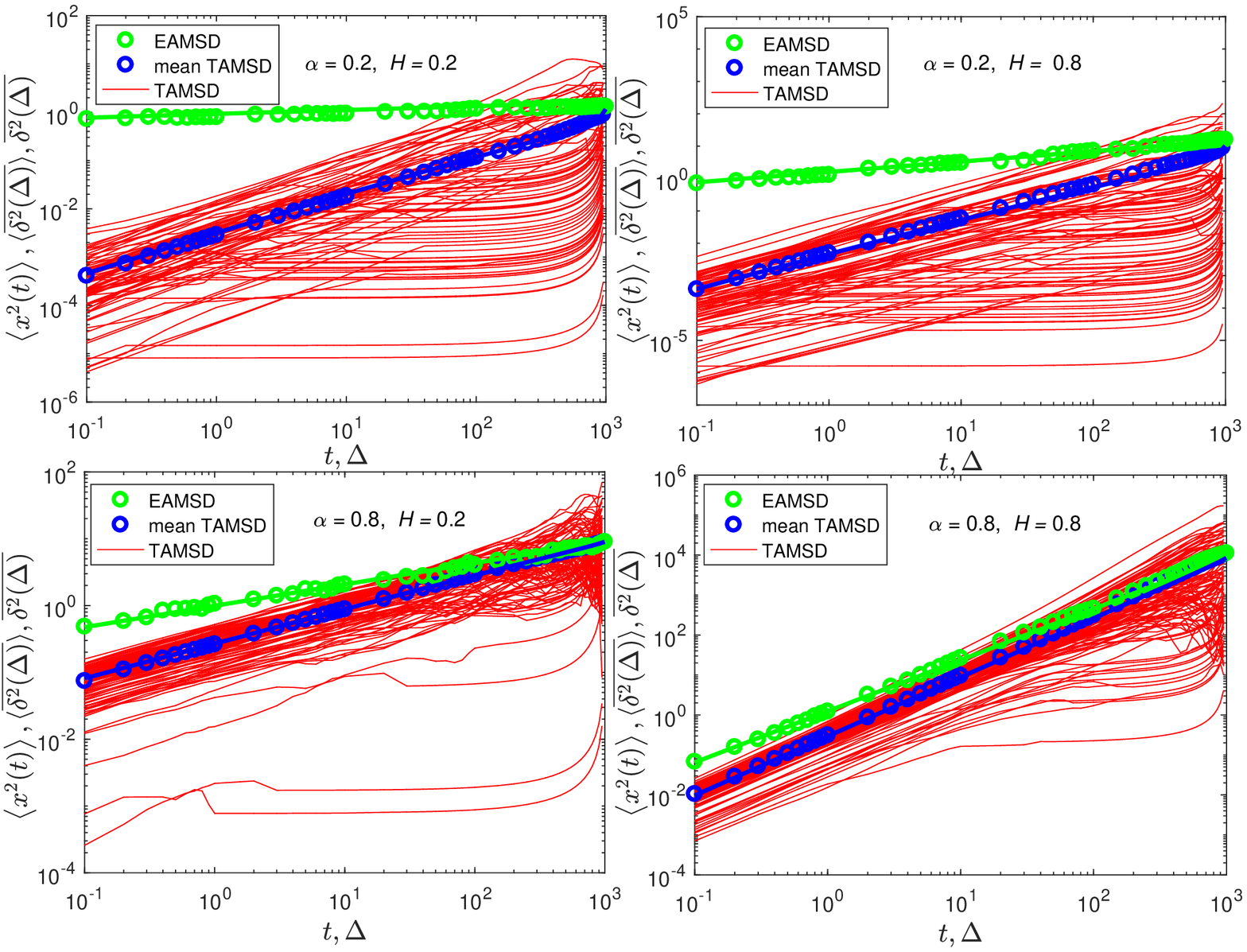}
\caption{MSD (green circles), individual TAMSDs (thin red curves), and mean
TAMSD (thick blue circles) for subordinated FBM without drift ($v=0$) for
different $H$ and $\alpha$. Theoretical results for the MSD (thick solid
green curves) and mean TAMSD (thick solid blue curves) are based on
Eqs.~\eqref{eq-7} and \eqref{eq-20}, respectively. Parameters: trajectory
length $T=1000$, elementary time-step $dt=0.1$, and number of trajectories
$n=300$.}
\label{fig-a2}
\end{figure}

\begin{figure}
\includegraphics[width=8cm]{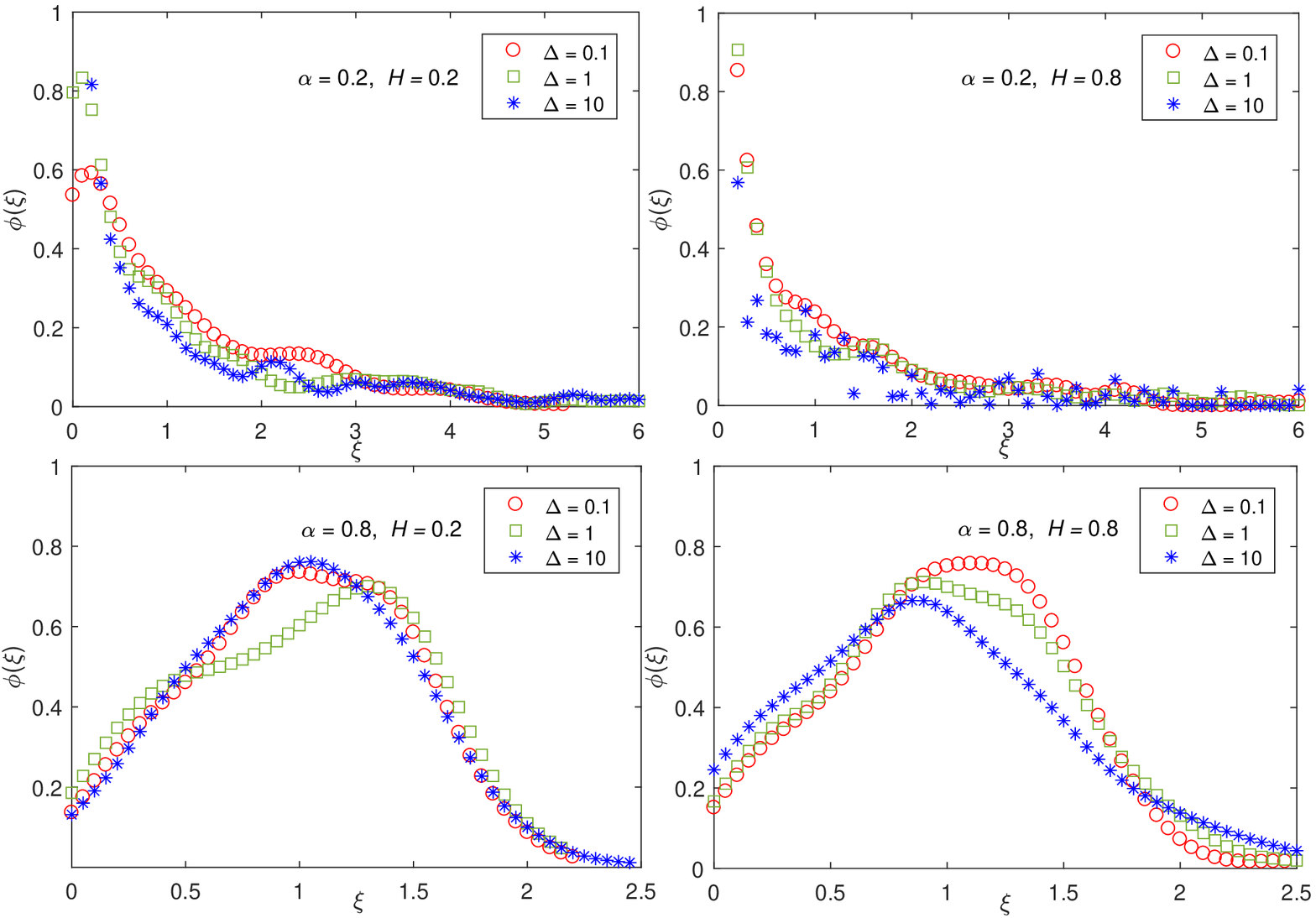}
\caption{Amplitude scatter PDF of TAMSDs for subordinated FBM without drift
($v=0$) for different $H$ and $\alpha$. The symbols (circle, square, and
star) respectively denote simulations with lag time $\Delta=0.1$, $1$, and
$10$.}
\label{fig-a3}
\end{figure}

\begin{figure}
\includegraphics[width=8cm]{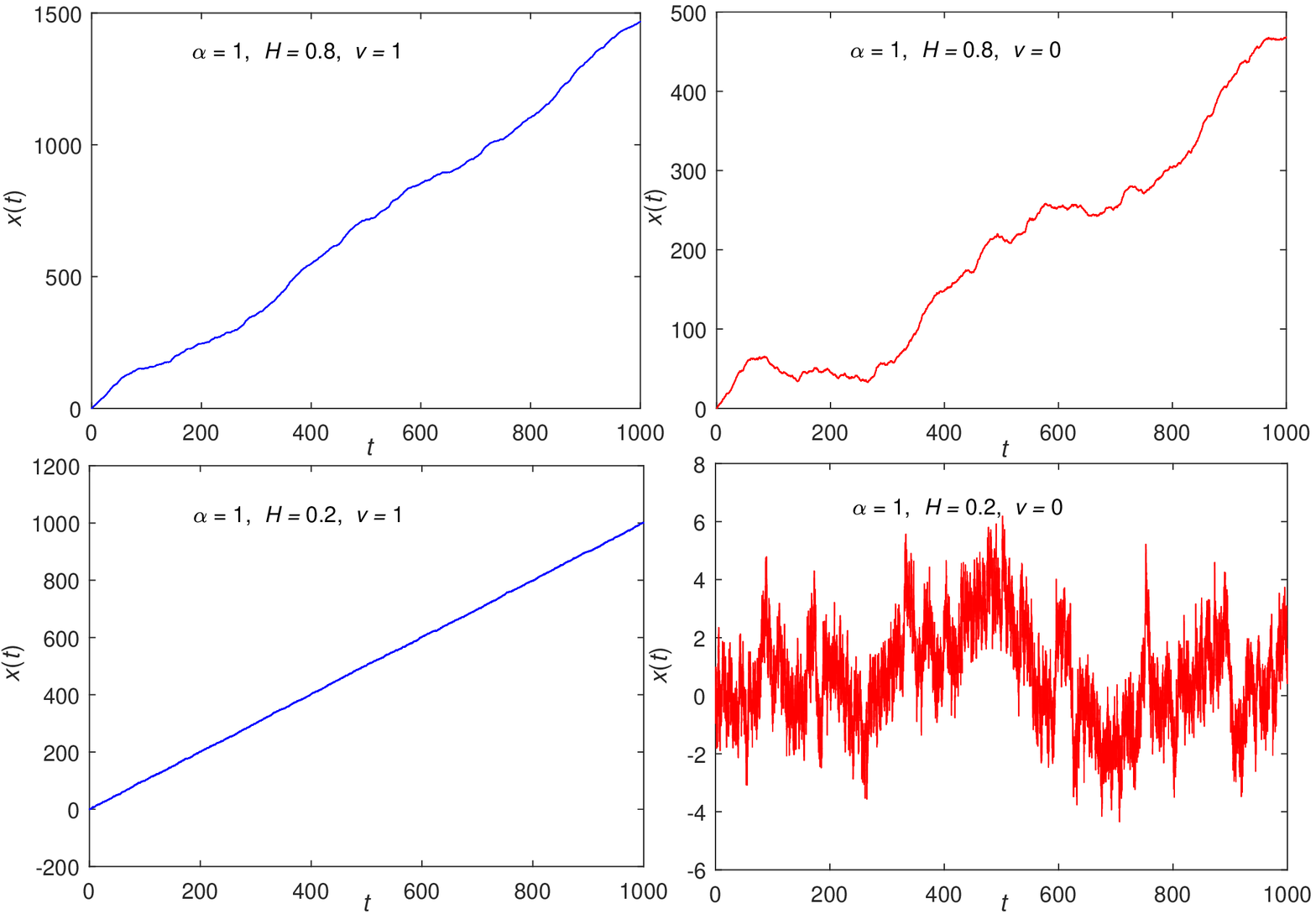}
\caption{Sample trajectories of subordinated FBM  with drift $v=1$ and
without drift ($v=0$) for the cases $H=0.8$ and $H=0.2$ with $\alpha=1$.}
\label{fig-a5}
\end{figure}

\begin{figure}
\includegraphics[width=8cm]{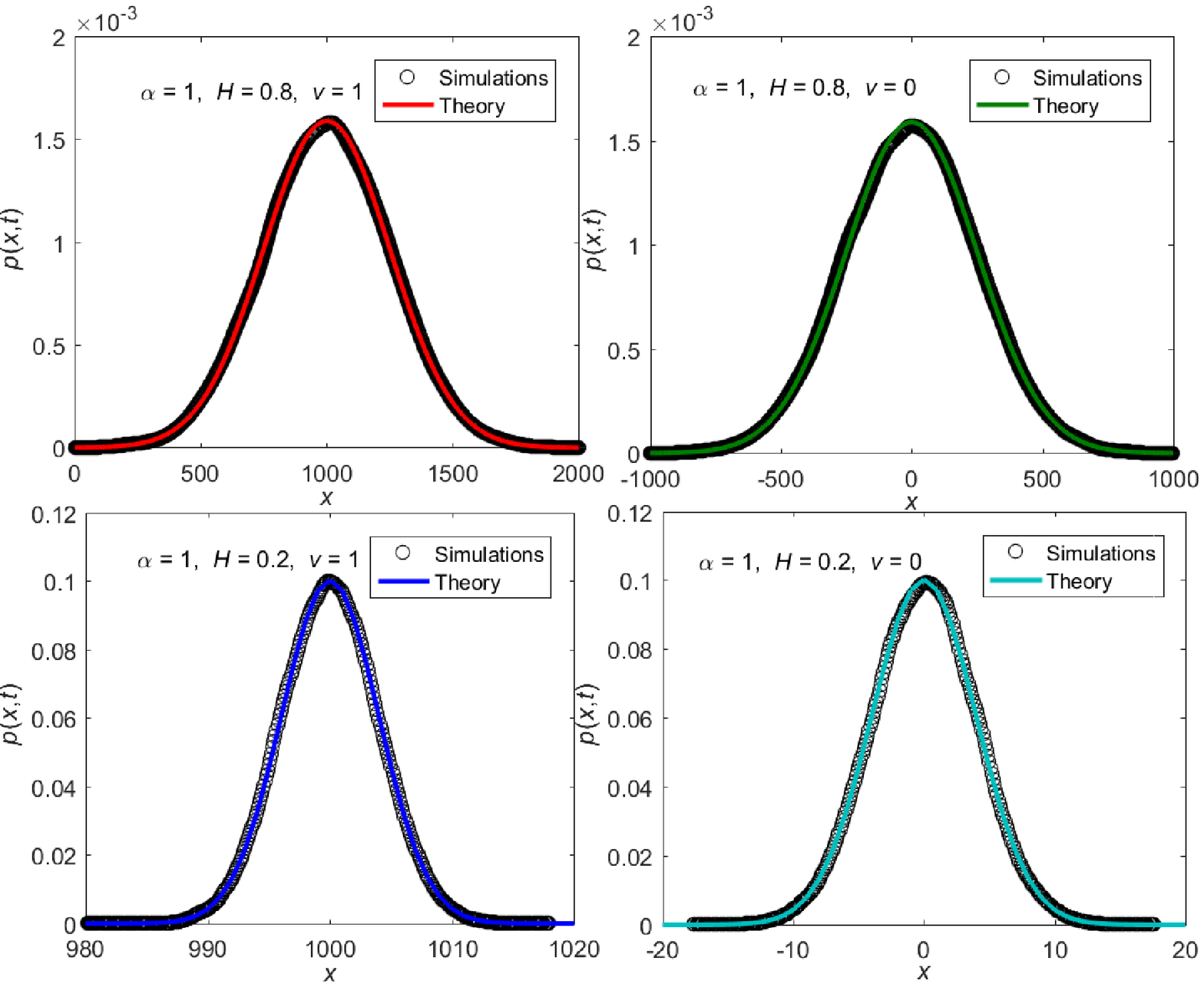}
\caption{Simulations (dark circles) and analytical results (solid curves)
from Eq.~\eqref{eq-4} for the PDF of subordinated FBM with drift $v=1$ and
without drift ($v=0$) for different $H$ and for $\alpha=1$ at $t=1000$.}
\label{fig-a6}
\end{figure}

\end{document}